\documentclass{aa}
\usepackage{graphics}
\newbox\grsign \setbox\grsign=\hbox{$>$} \newdimen\grdimen \grdimen=\ht\grsign
\newbox\simlessbox \newbox\simgreatbox \newbox\simpropbox
\setbox\simgreatbox=\hbox{\raise.5ex\hbox{$>$}\llap
     {\lower.5ex\hbox{$\sim$}}}\ht1=\grdimen\dp1=0pt
\setbox\simlessbox=\hbox{\raise.5ex\hbox{$<$}\llap
     {\lower.5ex\hbox{$\sim$}}}\ht2=\grdimen\dp2=0pt
\setbox\simpropbox=\hbox{\raise.5ex\hbox{$\propto$}\llap
     {\lower.5ex\hbox{$\sim$}}}\ht2=\grdimen\dp2=0pt
\def\simgreat{\mathrel{\copy\simgreatbox}}
\def\simless{\mathrel{\copy\simlessbox}}

\begin{document}
\thesaurus{08(08.01.1; 08.01.3; 08.02.1; 08.09.02 RE J0720-318)}
\title{Opacities along the Line of Sight to and in the Atmosphere of the White Dwarf in
the Close Detached DAO+dM Binary \object{RE J0720-318} (V$^{*}$ IN CMa).}
\author{P.D.Dobbie \inst{1}
 \and M.A.Barstow\inst{1}\thanks{Guest Observer with EUVE}
   \and M.R.Burleigh\inst{1}$^{\star}$
     \and I.Hubeny\inst{2}} 	
\institute{Dept. of Physics and Astronomy, University of Leicester, University Road, Leicester, LE1 7RH, UK 
\and NASA Goddard Space Flight Center, Code 681, Greenbelt, MD 20771} 
\offprints{P.D.Dobbie}
\mail{pdd@star.le.ac.uk}
\date{Received  / Accepted}
\titlerunning{The DAO White Dwarf \object{RE J0720-318}}
\authorrunning{P.D.Dobbie et al.}
\maketitle

\begin{abstract}
We present the results from a multi-wavelength study of the mixed H+He composition DAO
white dwarf \object{RE J0720-318}. A detailed analysis of UV and EUV spectroscopic 
data with state-of-the-art non-LTE photospheric models demonstrates that
the observed opacity to EUV radiation probably results from a more
complex structure than a simple
H+He, chemically layered atmosphere. Instead, EUV photometry
and phase resolved EUV spectroscopy indicate a likely spatial non-uniformity in
the surface distribution of helium, which is consistent with a model in which 
material is accreted from the wind of the dM secondary. The rotational
modulation of the spatially inhomogeneous EUV opacity allows us to estimate the rotation period 
of the white dwarf ($0.463 \pm 0.004$ days). We have also reviewed
 two plausible origins proposed by Burleigh et al. (\cite{burleigh97})
and Dupuis et al. (\cite{dupuis97a}) to account for the unusual N(HI)/N(HeI)$\sim$1 along this line of
sight. We conclude that
it is probably  due to the presence of a cloud of ionized gas
along this line of sight, rather than a
circumbinary disk. The cloud, residing between 123-170pc distant in the
direction of the CMa ISM tunnel, may be $\simgreat40$pc in length. 
\keywords{stars: abundances -- stars: atmospheres -- stars: binaries: close -- stars: individual: \object{RE J0720-318}}
\end{abstract}
\section{Introduction}
Soft X-ray and EUV observations of DA white dwarfs have revealed
photospheric opacity in excess of that provided by a pure hydrogen plasma for
the majority with  $T_\mathrm{eff} > 50\, 000\mathrm{K}$ (e.g. Marsh et
al. \cite{marsh97}; Wolff et al. \cite{wolff96}; Finley
\cite{finley96}; Jordan et al. \cite{jordan94}; Barstow et
al. \cite{barstow93}). In addition, high
resolution IUE echelle spectra of those stars exhibiting the greatest levels of
opacity show absorption lines of heavy elements (e.g. C, N, O, Si, Fe and
Ni; Bruhweiler \&
Kondo \cite{bruhweiler81}; Dupree \& Raymond \cite{dupree82}; Bruhweiler \&
Feibelman \cite{bruhweiler91}; Holberg et al. \cite{holberg94m}; Holberg, Barstow and Sion
\cite{holberg98a}). In many cases these have been shown to be
consistent with their location in the stellar 
photosphere (e.g. \object{G191-B2B}, Reid \&
Wegner \cite{reid88}). Early theoretical work had suggested that the atmospheres
of DAs were more or less pure hydrogen as heavier elements should rapidly
sink out in the strong gravitational field (Schatzmann
\cite{schatzmann58}). However, further detailed calculations have shown that
finite abundances of metals can be supported against gravitational
settling by the net transfer of upward momentum
from the intense radiation field to the heavy element ions via their line
transitions (e.g. Vauclair et al. \cite{vauclair79}
and most recently
Chayer
 et al. \cite{chayer95}). In contrast, similar calculations have
demonstrated that helium cannot be radiatively levitated in observable
quantities in all but the
hottest objects, due to a relatively small number of bound-bound
transitions, and should sink out of the photosphere on a time-scale
of days (Vennes et al. \cite{vennes88}).

White dwarfs are classified according to their optical spectra
and generally fall into one of two categories. The hydrogen rich DAs
and the helium rich DOs display solely strong H and He lines respectively. 
The DAOs are a hybrid class and show both strong H
lines and a weak 4686\AA\ HeII feature. Lying in the temperature range
$50\,000 - 70\,000 \mathrm{K}$, Fontaine \& Wesemael (\cite{fontaine87}) proposed that
 they were transitional
objects forming an evolutionary link between the DOs and DAs. As a DO
cooled, residual hydrogen in its atmosphere was expected to float to the surface
forming a layer of sufficient thickness that spectroscopic traces of
helium were eliminated, at least in the optical waveband.  The white dwarf
would transform first into a DAO, then a DA. In either case it was
expected to have a chemically stratified atmosphere where a thin layer 
of hydrogen over-lies
a predominantly helium envelope in diffusive equilibrium. At $30\,000 \mathrm{K}$, convective mixing
would then dredge helium back into the optical photosphere, effectively
returning it to the helium rich cooling sequence. Many
observations could be accounted for with this evolutionary hypothesis, 
e.g. the apparent absence of DAs with $T_\mathrm{eff} >
 70\,000 \mathrm{K}$ and the DO/DB gap between
$30\,000 - 45\,000 \mathrm{K}$, a region of the cooling sequence where no helium
rich objects are observed. However, doubt was cast on this interpretation by the
detailed optical study of a large sample of DAOs by Bergeron
et al. (\cite{bergeron94}). In contrast to the theoretically predicted 
stratified atmosphere, they found for many objects in their sample that the observed 
4686\AA\ HeII line profile was reproduced significantly
better by a model in which helium was homogeneously mixed
into a predominantly hydrogen atmosphere. In addition, the low surface
gravities of many DAOs suggested that they were products of extreme
horizontal branch evolution (EHB), inconsistent
 with the generally accepted post-AGB evolutionary
status of the DOs and DAs (e.g. Weidemann \cite{weidemann90}).

\object{RE J0720-318} is one of only three DAOs detected in the
EUV sky surveys of the ROSAT Wide Field Camera (WFC, Pounds et
al. \cite{pounds93}) and the Extreme Ultraviolet Explorer (EUVE, Bowyer et
al. \cite{bowyer94}). Optical follow up observations revealed a post
common envelope (CE) DAO+dM binary with a period of 1.26 days
 (Vennes \& Thorstensen \cite{vennes94a};  Barstow et al. 
\cite{barstow95}; \ Vennes \& Thorstensen
\cite{vennes96a}). Intriguingly all three of the DAOs detected in the EUV
reside in close, pre-cataclysmic variable (pre-CV) systems and have
masses typical of isolated DAs (e.g. Bergeron et
al. \cite{bergeron92}) and consistent with post-AGB
evolution. Radiative levitation has been shown to be insufficient to account
for the observed abundance of helium in these
objects (Vennes et al. \cite{vennes88}) and it is thought binarity is most likely
responsible (Tweedy et al.
\cite{tweedy93}). It is feasible that tidal effects,
accretion from the secondary's wind and the earlier period of common envelope
evolution may all have played or be playing a role in determining the
observed abundance patterns (Bergeron et al. \cite{bergeron94}).
Current observational evidence seems to favour a wind accretion
mechanism. For example, periodic variations in the soft X-ray/EUV
flux from the DA2+KV pre-CV system \object{V471 Tauri} (Jensen et
al. \cite{jensen86}) can be
consistent with the rotational modulation of regions of increased
helium and metal opacity on the surface of the white
dwarf (Dupuis et al \cite{dupuis97b}). Recently, time resolved studies
of a Zeeman split 1286\AA\ SiIII line believed to be formed in the
photosphere of the \object{V471 Tauri} white dwarf have revealed a variation in
its equivalent width on the
soft X-ray/EUV and optical period (Sion et al. \cite{sion98}). The
time of greatest
line strength corresponds to soft X-ray/EUV minimum and optical
maximum, more or less confirming the magnetic wind
accretion interpretation first proposed by
Jensen et al. (\cite{jensen86}). The influence of tidal forces, the CE
phase and weak mass-loss is less clear and observations are essential for constraining theoretical models of close
binary evolution. In an initial analysis of the EUVE spectrum of \object{RE J0720-318}, Burleigh et al. (\cite{burleigh97}) 
were able to reproduce the data with a simple LTE H+He
stratified model. The unexpectedly
low H-layer mass of $3 \cdot 10^{-14}$M$_{\sun}$ led them to speculate
that the white dwarf may have been largely stripped of hydrogen by
mass transfer during the prior common envelope phase. Subsequently a weak wind was
proposed to be carrying helium up into the photosphere or alternatively accretion from the
secondary's wind depositing helium into the optical line forming
region to produce the 4686\AA\ HeII feature.

We report here the results of a new multi-wavelength study of \object{RE J0720-318}. Using state-of-the-art non-LTE models we are able to
demonstrate that the opacity observed in the EUV waveband probably
arises from a more complex structure than a simple, LTE H+He stratified atmosphere. Using EUV photometry and
phase resolved spectroscopy we also show that it is likely the helium abundance is not
spatially uniform across the photosphere, in support of a wind
accretion model. As a result we have been able to measure the rotation
rate of the white dwarf ($0.463 \pm 0.004$ days).

%__________________________________________________________________

\section{Observations and data reduction.}
\subsection{The EUVE data}

An EUVE observation of \object{RE J0720-318} was carried out during
the period 1995 December 16--20 and is now available from the public
archive at Goddard Space Flight Centre. The target was
observed in dither mode, which consists of a series of discrete
pointings slightly offset from each other and is carried out to
minimize the effects on the data of detector spatial response variations.
 \object{RE J0720-318} was detected in all three spectrometers and
by the Deep Survey (DS) imager with
effective exposure times of 122\,459s \ (SW 70--190\AA), 121\,067s \ (MW
140--380\AA), 112\,861s \ (LW 280--760\AA) and 112\,564s \ (DS 70--180\AA).
 Standard IRAF EUV and XRAY/PROS software were used to
extract light-curves of \object{RE J0720-318} from each of the four detectors,
following procedures described in detail on the CEA website at
Berkeley\footnote{http://www.cea.berkeley.edu}. The count rates for all four were corrected for
detector dead-time and an effect known as ``Primbsching'' which occurs when
the satellite telemetry buffer saturates, resulting in some photon events being lost.
 Additionally, the DS count
rates were modified to account for times when the source lay in the
deadspot, a region of the detector with reduced gain (e.g. Sirk et
al. \cite{sirk97}). The spectral extractions were also performed using
IRAF/EUV software, following our standard procedures for reducing
spectroscopic data which includes quadratically adding a 5\%\ systematic error to
the data to account for residual
detector fixed pattern efficiency variations (FPN). A more detailed
description of these methods may be found in our earlier
work (e.g. Barstow et al. \cite{barstow97}).

\subsection{The HST GHRS and optical observations.}
A total of six HST GHRS observations of \object{RE J0720-318} were obtained from the
public archive at STSCI. These were
taken with the G160M grating through the Large Science Aperture (LSA)
and cover a waveband of 35\AA\ with a resolution better than
0.1\AA. Each dataset consists of several shorter exposures which were co-added to
 produce the resulting spectra. A Pt/Ne hollow cathode
lamp exposure (WAVE) was acquired immediately prior to each of the
science observations, allowing us to use IRAF/STSDAS
routines to re-calibrate the wavelength solution and subsequently
reduce the uncertainty in the wavelength scale from $15 \mathrm{km s}^{-1}$ to
$4 \mathrm{km s}^{-1}$. The observations were timed to coincide with binary
quadrature as predicted by the optically derived ephemeris of Vennes
\& Thorstensen (\cite{vennes96a}). Details of these, including
wavebands, exposure times and the predicted binary phase are shown in Table~\ref{table1}.
\begin{table}
\caption[The HST Observation Log.]{The HST observation log -- the time
and the date given are
for mid-exposure; phase is calculated from the Vennes
\& Thorstensen (\cite{vennes96a}) ephemeris, assuming P = 1.26245 days and
T$_{0}$(HJD) = 2449735.327. }
\label{table1}
\scriptsize
\begin{center}
\begin{tabular}{cccccc} \hline
Obs. ID.&Date&Time(UT)&$\lambda$(\AA)&Exp.(s)&$\Phi$ \\ \hline
z3c80405 & 17/09/96 & 00:06 & 1623--1658 & 3x571 & 0.74 \\ 
z3c80408 & 17/09/96 & 01:35 & 1623--1658 & 4x598 & 0.79 \\
z3c80305 & 25/10/96 & 11:08 & 1532--1568 & 3x571 & 0.21 \\
z3c80308 & 25/10/96 & 12:38 & 1532--1568 & 4x598 & 0.26 \\
z3c8030b & 25/10/96 & 14:15 & 1623--1658 & 4x598 & 0.31 \\
z3c8030e & 25/10/96 & 15:51 & 1623--1658 & 4x598 & 0.36 \\ \hline 
\end{tabular}
\end{center}
\end{table}
\normalsize
The analysis of UV and EUV data is complemented by a re-analysis of a
co-added optical spectrum of \object{RE J0720-318} obtained on the 1.9m
Radcliffe telescope at SAAO in February 1995. Full details of this
dataset may be found in Barstow et al. (\cite{barstow95}).
 
\section{Analysis and results.}
\subsection{Models and spectral analysis techniques.}
Our non-LTE photospheric models are generated with the latest version of
 the atmosphere code TLUSTY
(version 195, Hubeny \cite{hubeny88}; Hubeny and Lanz \cite{hubeny95}) and 
the spectral synthesis program SYNSPEC (Hubeny
et al. \cite{hubeny94}) and include the full effects of
line blanketing. Detailed hydrogen and helium line profiles are calculated from
the tables of Sch$\rm \ddot{o}$ning
 \& Butler  (private comm.) and Sch$\rm \ddot{o}$ning
 \& Butler (\cite{schoning89}) respectively. 
To account for the opacity to EUV radiation of HI, HeI and HeII along
the line of sight, during the analysis of EUV spectroscopic data,
 photospheric models are used in conjunction with the ISM
model of Rumph et al. (\cite{rumph94}). This has been modified to
include the effects of the converging Lyman line series of HeI and HeII at
504\AA\ and 228\AA\ respectively. 

Our standard spectral analysis techniques have been described 
in detail in previous publications
(e.g. Barstow et al. \cite{barstow97}) and will only be briefly
re-iterated here.  We carry out comparisons between models and
data using the spectral fitting program XSPEC (Shafer et al. \cite{shafer91}). XSPEC
works by
folding a model through the instrument response before comparing the
result to the
data by means of a $\chi^{2}$-statistic. The best fit model
representation of the data is found by incrementing free grid
parameters in small steps, linearly interpolating between points in
the grid,
until the value of $\chi^{2}$ is minimized. Errors are  
calculated by stepping the parameter in question away from its optimum value
and re-determining the minimum $\chi^{2}$. This is repeated until the difference between the
two values, $\Delta\chi^{2}$, corresponds to 1$\sigma$ for a given
number of free model parameters as prescribed by Lampton et al. (\cite{lampton76}). All errors quoted in this work are
1$\sigma$ unless stated otherwise.

\subsection{The optical data analysis}
Previous studies of the optical spectrum of \object{RE J0720-318} have employed
LTE H+He models (e.g. Barstow et al. \cite{barstow95}). A re-analysis
of the optical spectrum has been carried out with a grid of non-LTE,
homogeneous, H+He models, primarily for consistency, since non-LTE
models are used here in the analysis of both the UV and EUV
data. Recently it has been demonstrated that
estimates of the effective temperatures of hot DAs determined by
fitting non-LTE model Balmer lines to the observed profiles are
systematically lower by $\approx 4000 \mathrm{K}$ at $\approx 60\,000 \mathrm{K}$ compared
to the earlier LTE analyses (Barstow et al. \cite{barstow98}). Barstow et
al. (\cite{barstow98}) also report that in
contrast to the results of Bergeron et al. (\cite{bergeron94}) who
found that the inclusion of small amounts of helium into LTE models lowers the
determined effective temperature by several $1000 \mathrm{K}$ at
$\approx 60\,000 \mathrm{K}$
(Bergeron et
al. \cite{bergeron94}), no similar effect occurs in
non-LTE. Fitting our grid of non-LTE homogeneous H+He models we determine
$T_\mathrm{eff} = 56\,600^{+1740}_{-1690}\mathrm{K}$,
 $\log \mathrm{g} = 7.41 \pm 0.1$ and a helium
abundance of $\log \mathrm{(He/H)} = -3.26^{+0.22}_{-0.23}$.
 We adopt $T_\mathrm{eff} = 55\,000 \mathrm{K}$ and $\log \mathrm{g} = 7.5$ for 
the subsequent analysis of UV and EUV data.
\subsection{The GHRS spectra}
\begin{figure*}
\resizebox{\hsize}{!}{\includegraphics{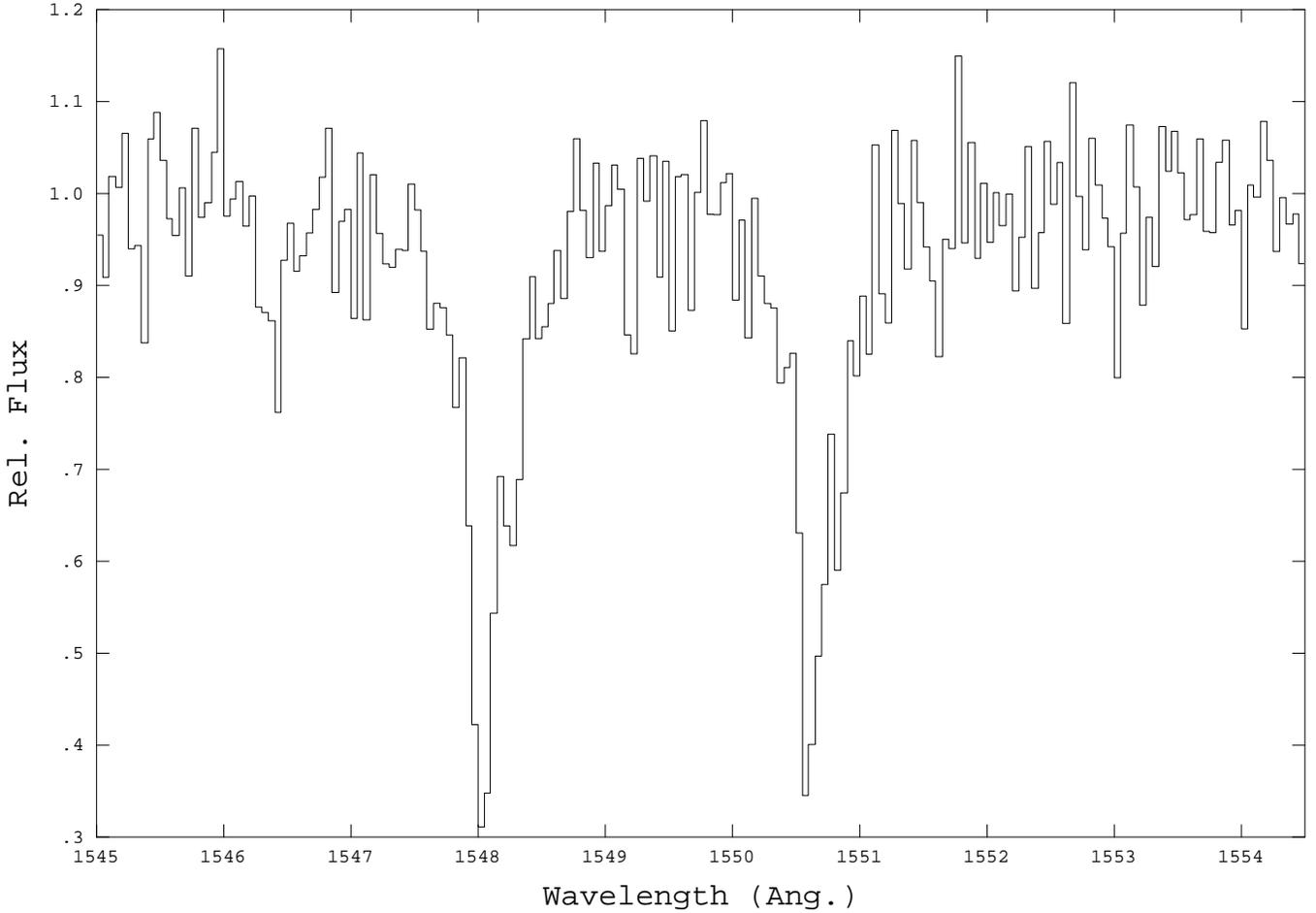}}
\caption{The spectrum z3c80308 covering the CIV doublet. The line profiles
may be complex and display weaker absorption redward of the main components.}
\label{carbon}
\end{figure*}
\begin{figure*}
\resizebox{\hsize}{!}{\includegraphics{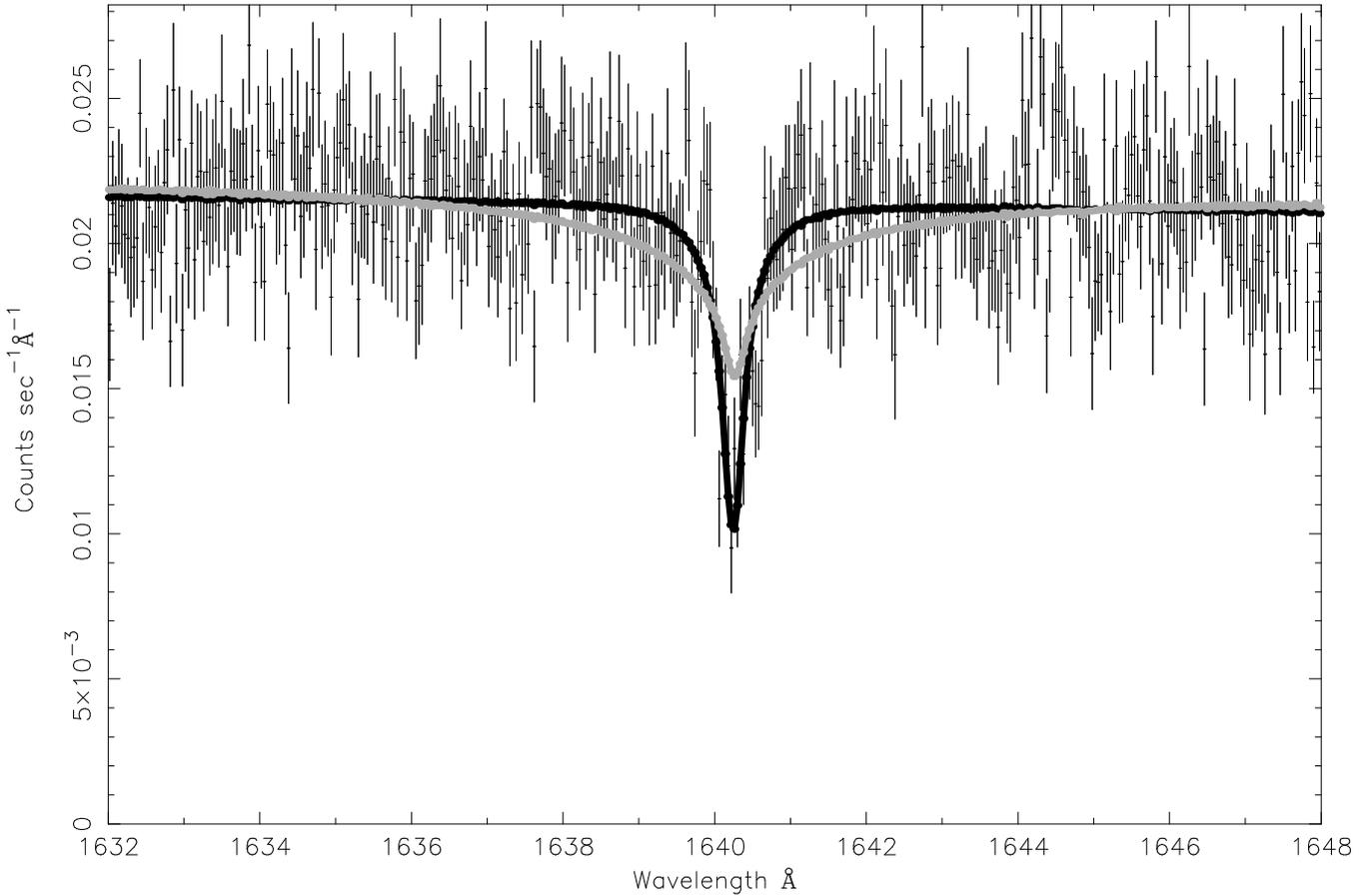}}
\caption{The best fit homogeneous (black) and stratified (grey) models to the GHRS
spectrum z3c8030b. See Table~\ref{table3} for details.}
\label{UVline}
\end{figure*}
\begin{figure*}
\resizebox{\hsize}{!}{\includegraphics{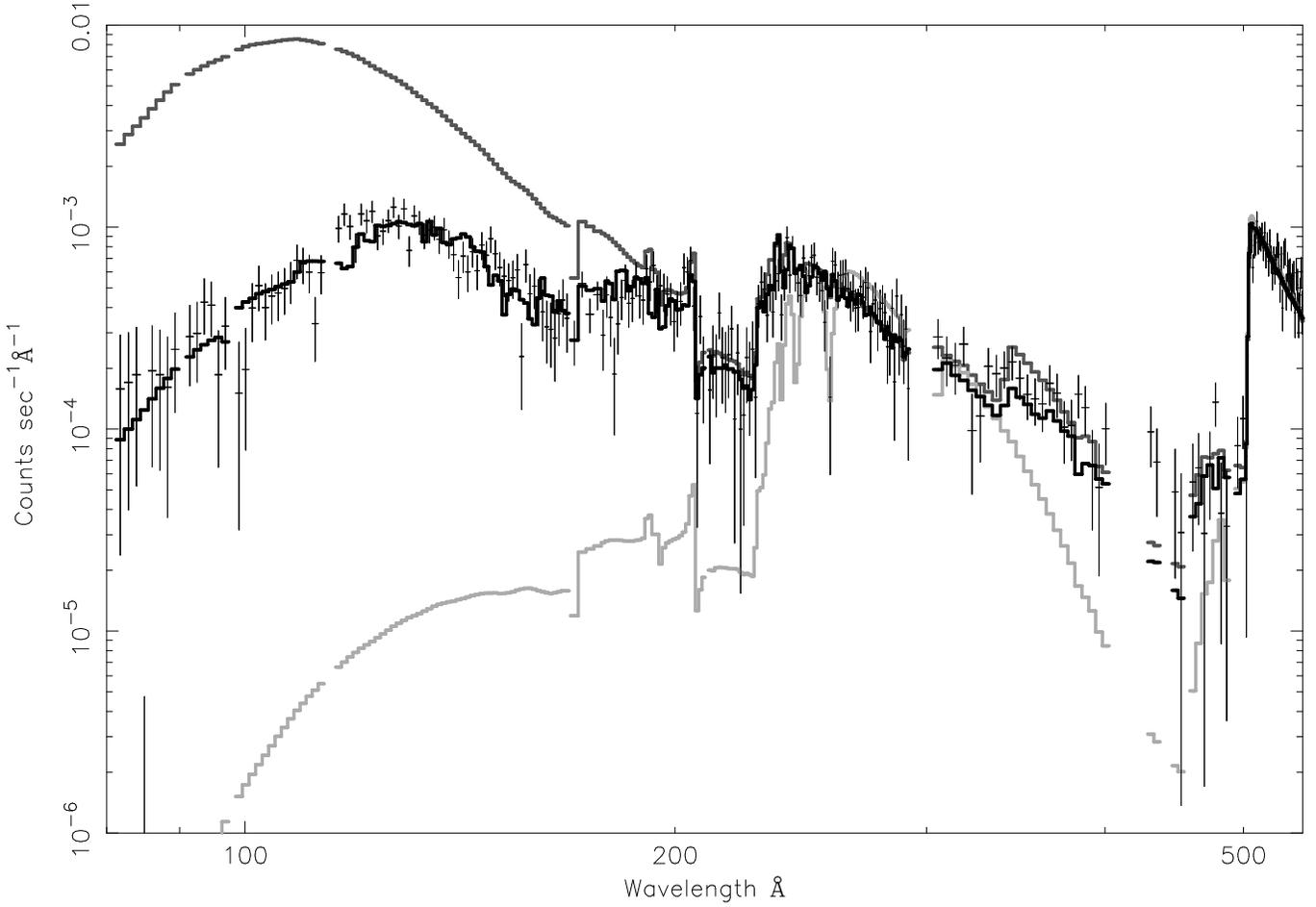}}
\caption{The non-LTE model fits to the EUV spectrum of \object{RE
J0720-318}, {\bf a.} the homogeneous
H+He+C (dark grey), N(HI) $= 2.5\cdot10^{18}\mathrm{cm}^{-2}$, N(HeI) $= 1.5\cdot10^{18}\mathrm{cm}^{-2}$,
N(HeII) $= 7.5\cdot10^{17}\mathrm{cm}^{-2}$, log (He/H) = -3.8, log
(C/H) = -5.5, {\bf b.} the
stratified H+He+C (light grey), N(HI) $= 2.5\cdot10^{18}\mathrm{cm}^{-2}$, N(HeI) $= 1.6\cdot10^{18}\mathrm{cm}^{2}$,
N(HeII) $= 5.1\cdot10^{12}\mathrm{cm}^{-2}$, log (H-layer) = -14.15, log (C/H) = -5.6 and {\bf c.}
the homogeneous models incorporating additional heavy elements
(black), see Table~\ref{table4}.} 
\label{euvehs}
\end{figure*}
\begin{figure*}
\resizebox{\hsize}{!}{\includegraphics{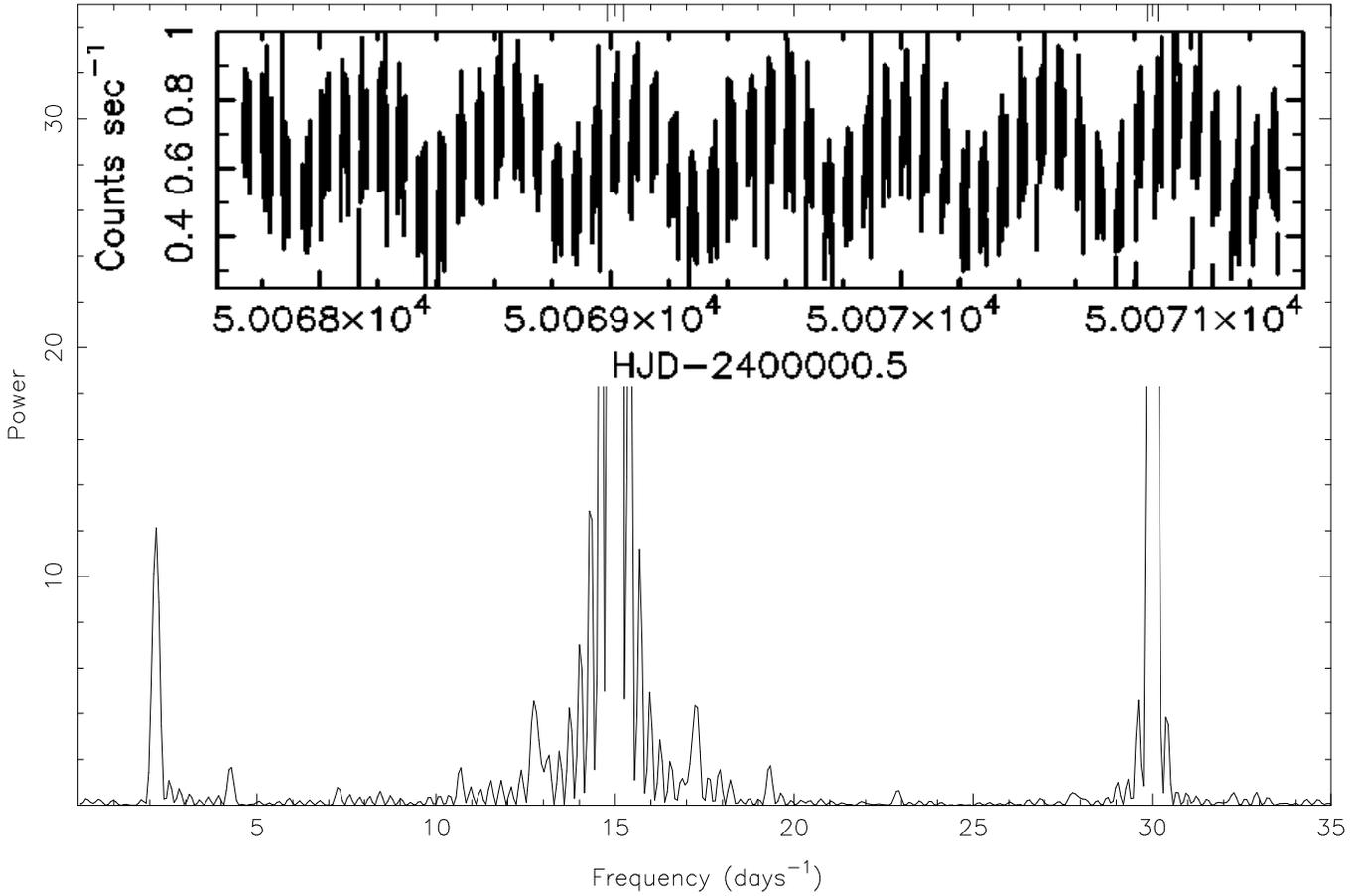}}
\caption{Inset: The DS light-curve. Main: Power spectrum of DS light curve.
 Power at 14.984 days$^{-1}$
and 30.007 days$^{-1}$ can be attributed to the satellite orbit. The
peak at 2.160 days$^{-1}$ is significant at the 3$\sigma$ confidence
level. }
\label{Period search}
\end{figure*}
\subsubsection{Line velocity measurements}

The two spectra covering the 1532-1568\AA\ range both display a prominent CIV resonance doublet. A glance at Fig.~\ref{carbon} indicates
that the doublet profiles may be complex with weaker, red-shifted
absorption lines present besides the main components. A test of the significance of these was carried out by applying an F-test to the fit statistics obtained using XSPEC to match both single and double gaussian profiles to the observed lines. The result  of this suggests they are real features ($>99.999$\%\ confidence).  A 1640\AA\ HeII absorption line is displayed in the four spectra covering the 1623--1658\AA\ waveband. This feature was assumed to
arise in the stellar atmosphere as photospheric helium
 has previously been detected in both the optical (Barstow et
al. \cite{barstow95}; Vennes \& Thorstensen \cite{vennes94a}) and the 
EUV spectrum
(Burleigh et al. \cite{burleigh97}; Dupuis et al.\cite{dupuis97a}) of
  \object{RE J0720-318}. Although the four spectra covering this waveband are
rather noisy, close examination of z3c8030b suggests that
there may also be a weak redward component to this profile, which can
be consistent in velocity space with the weaker CIV features.
However, from fitting gaussains to similar looking features in this
spectrum which are attributable to noise, it was found that these
could also be identified as absorption lines at a similar level of confidence ($<99$\%).  Furthermore, the spectrum taken immediately after and covering the same waveband (z3c8030e), does not appear to display this feature. Subsequently, the velocities of the lines in the six spectra were measured using the STARLINK package DIPSO to fit gaussians or where appropriate, double gaussians, to the observed profiles. The errors were calculated by quadratically
 adding the uncertainties in both the fitting
procedure and in the wavelength calibration.  The heliocentrically correct velocities along with their associated errors are given in Table~\ref{table2}.

The HeII velocities were used in conjunction with the binary ephemeris
of Vennes \& Thorstensen (\cite{vennes96b})  to constrain the origin of features
observed in the 1532-1568\AA\ waveband. Unfortunately, \ since Vennes \& Thorstensen (\cite{vennes96a}) do not give their measured 4686\AA\ HeII line
velocities in tabular form, it has not been possible to fully update
this ephemeris. Instead, the HeII measurements were used to adjust
K$_{\mathrm{WD}}$ and $\gamma_{tot}$, the white dwarf semi-amplitude
and the velocity offset (gravitational and system) respectively. Adopting the Vennes \& Thorstensen binary
period of P = 1.26245 days and epoch(HJD) of T$_{0} = 2449735.327$,
both of which should be sufficiently accurate for the purposes here,
sinusoids of the form V(T) = $\gamma_{tot}$ + K$_{\mathrm{WD}}$ $\sin$
[$\pi$ + $2\pi$(T - T$_{0}$) / P] were fit to the four data
points. The best match was found between the observed and predicted
velocities for a value of K$_{\mathrm{WD}} = 76.2 \pm 5.9
\mathrm{kms}^{-1}$ and $\gamma_{tot} = 38.4 \pm 6.0
\mathrm{kms}^{-1}$. Comparing these figures to the Vennes \&
Thorstensen (\cite{vennes96a}) values of K$_{\mathrm{WD}} = 114 \pm 13
\mathrm{kms}^{-1}$ and $\gamma_{tot} = 51 \pm 10 \mathrm{kms}^{-1}$,
within their respective errors, consistency is found between estimates of $\gamma_{tot}$. 
However, the difference in K$_{\mathrm{WD}}$ is rather large.  It is
possible that the weak 4686\AA\ HeII profile from which their values for
K$_{\mathrm{WD}}$ and $\gamma_{tot}$ were determined, may be contaminated with 
emission from the irradiated face of the
secondary star. It is interesting to note from Fig.~2 of Vennes \&
Thorstensen (\cite{vennes96a})  that the
two velocities which
agree best with the lower value for K$_{\mathrm{WD}}$ are obtained at phases
0.783 and 0.828 where re-processed emission is approaching a minimum
and the HeII profile appears deepest. Furthermore, Vennes \&
Thorstensen (\cite{vennes96a})  
point out that the errors on their velocity values are
formal and are likely to underestimate the true uncertainties.
Comparing the measured CIV line velocities to the predictions of
the modified ephemeris, it was found that the
main components are consistent with the predicted photospheric
velocity of the white dwarf. If this interpretation is correct then this
represents the first direct detection of carbon in the atmosphere of this white
dwarf. The weaker components, are red-shifted by
$\sim 50 \mathrm{km s}^{-1}$ with respect to the apparent photospheric
velocity. We note that the system velocity is $28 \pm 3 \mathrm{km s}^{-1}$ (Vennes \&
Thorstensen \cite{vennes96a}). The possible origin of these features will be 
discussed later.
%%%%%%%%%%%%%%%%%%%%%%%%%%%%%%%%%%%%%%%%%%%%%%%%%%%%%%%%%%%%%%%%%%
\begin{table}
\caption[UV Line Velocities]{The observed and laboratory wavelengths
for the lines detected in each of the GHRS spectra. Top, the lines
interpreted as photospheric and lower the possible weak red-shifted CIV
components. The heliocentric velocities of these lines are also shown.}
\label{table2}
\scriptsize
\begin{center}
\begin{tabular}{llccr} 
\hline 
Obs. ID. & Ion & Lab. $\lambda$ & Obs. $\lambda$ & Vel.\\
&&  (\AA)&  (\AA)&(km s$^{-1}$)  \\ \hline
405 & HeII & 1640.474 & $1641.099\pm0.030$ & $ 114.2\pm6.8$    \\
408 & HeII & 1640.474 & $1641.088\pm0.048$ &  $112.2\pm9.7$  \\
305 & CIV  & 1548.195 & $1548.052\pm0.043$ &  $-27.7\pm9.3$    \\
305 & CIV  & 1550.770 & $1550.606\pm0.015$ & $-31.7\pm5.0$   \\
308 & CIV &  1548.195 & $1548.027\pm0.014$ & $-32.5\pm4.9$   \\
308 & CIV &  1550.770 & $1550.606\pm0.030$ & $-31.7\pm7.1$   \\
30b & HeII & 1640.474 & $1640.286\pm0.025$ & $-34.6\pm6.1$   \\
30e & HeII & 1640.474 & $1640.379\pm0.030$ &  $-17.4\pm6.9$   \\ \hline
305 & CIV  & 1548.195 & $1548.282\pm0.015$ & $16.8\pm5.0$   \\
305 & CIV  & 1550.770 & $1550.854\pm0.033$ & $16.2\pm4.8$   \\
308 & CIV &  1548.195 & $1548.285\pm0.026$ & $17.4\pm6.5$  \\
308 & CIV &  1550.770 & $1550.871\pm0.030$ & $19.5\pm7.6$  \\ \hline
\end{tabular}
\end{center}
\end{table}
%%%%%%%%%%%%%%%%%%%%%%%%%%%%%%%%%%%%%%%%%%%%%%%%%%%%%%%%%%%%%%%%%%%%
\subsection{The abundances of He and C from UV data}

Two small grids of models were generated to measure the photospheric
abundances of helium and carbon. The first grid consists of
models including a homogeneous mix of H, He and C. The second grid consists of models with a stratified
atmospheric structure where the distribution of helium in the predominantly
hydrogen envelope was calculated by integrating the equations of
ordinary diffusion for a given hydrogen layer mass (Vennes et al.
\cite{vennes88}).  The homogeneous model grid was specified by the
parameters log (He/H) and log (C/H) and the stratified grid by the parameters
log(H-layer mass) and log (C/H). The analysis of the UV spectra was carried out in a three stage
procedure which was repeated for each grid. First the carbon abundance was
measured allowing the helium abundance/hydrogen layer mass to settle
at an arbitrary value between the limits of the calculated grid. Second, with the carbon abundance frozen at
the average value from the first stage, the grid was fit to the
four spectra displaying the helium absorption profile. In the final
step, with the helium abundance/hydrogen layer mass held at the average
value obtained in the previous step, the carbon abundance was
re-measured. The results from the two latter stages are given in Table~\ref{table3}. The best fitting homogeneous and stratified models to
spectrum z3c8030b are shown in Fig.~\ref{UVline}. It is clear that
while the stratified structure predicts a HeII line core which is too
shallow, the homogeneous model with a helium
abundance of $\log \mathrm{(He/H)} \approx -3.74$ produces a good
match. Similarly, in the remaining three spectra covering
 this wavelength range, the shape of the HeII
1640\AA\ line is only accurately reproduced by a homogeneous
model.
%%%%%%%%%%%%%%%%%%%%
%table of abundances
%%%%%%%%%%%%%%%%%%%%
\begin{table}
\caption[UV abundances]{The best fitting homogeneous model abundances to the
photospheric lines observed in the GHRS spectra. Also given, where
applicable are the best fit H-layer masses for the stratified model
structure. (* limit of calculated model grid).
}
\label{table3}
\footnotesize
\begin{center}
\begin{tabular}{cccl} \hline
Obs. ID.& Element & Abundance & H layer mass \\
& & log (N/H) & log (M$_{\sun}$) \\ \hline
z3c80405 & He & $-4.00^{+0.25}_{-0.25}$ & $-14.30^{+0.15}_{-0.14}$\\ 
z3c80408 & He & $-4.02^{+0.23}_{-0.22}$ & $-14.29^{+0.13}_{-0.13}$ \\
z3c80305 & C & $-5.46^{+0.15}_{-0.15}$ & not applicable \\
z3c80308 & C & $-5.28^{+0.12}_{-0.14}$ & not applicable \\
z3c8030b & He & $-3.74^{+0.20}_{-0.24}$ & $-14.42^{+0.13}_{-0.08*}$ \\
z3c8030e & He & $-4.04^{+0.22}_{-0.21}$ & $-14.36^{+0.14}_{-0.13}$ \\ \hline 
\end{tabular}
\end{center}
\end{table}
%%%%%%%%%%%%%%%%%%%%
\subsection{The EUVE data}
\subsubsection{The EUV spectrum}
The grids of models calculated for the analysis of the UV spectra were
extended to cover the EUV wavelength range. During the fitting
process, the synthetic spectra were normalized to a V magnitude
point and the ISM model parameters HI, HeI and HeII allowed to vary
freely. The hydrogen layer/helium abundance and
carbon abundance grid parameters were constrained
 by the most extreme 1$\sigma$ limits determined in the analysis of
the UV data.  An adequate fit could not be obtained with either
model. As shown in Fig.~\ref{euvehs} the best fit stratified
structure predicted a flux between 80--240\AA\ several orders of magnitude
lower than observed, effectively ruling out this
layered atmospheric configuration. On the other hand, although the homogeneous
model could provide a good match to the observed spectrum at
wavelengths above 180\AA, shortward of this the predicted flux is too
high. 
However, after the inclusion of other heavy elements including N,O,Si,Ne and Fe in the abundances shown in Table~\ref{table4}, 
an excellent match between the synthetic spectrum and data was
achieved. Given that it was necessary to include these elements, this
analysis strongly suggests 
that other metals, besides carbon, are present in the atmosphere of
\object{RE J0720-318}. We note that Dupuis et al. (\cite{dupuis97a})
have previously fit the spectrum with a homogeneous, LTE model
including trace heavy elements.
\subsubsection{The EUV Light Curve of \object{RE J0720-318}.}

\begin{figure}
\resizebox{\hsize}{!}{\includegraphics{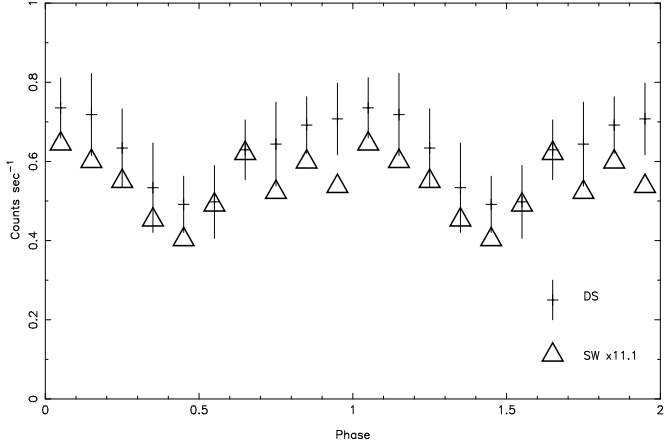}}
\caption{DS and SW light curves folded on our ephemeris.}
\label{Folded}
\end{figure}

\begin{figure*}
\resizebox{\hsize}{!}{\includegraphics{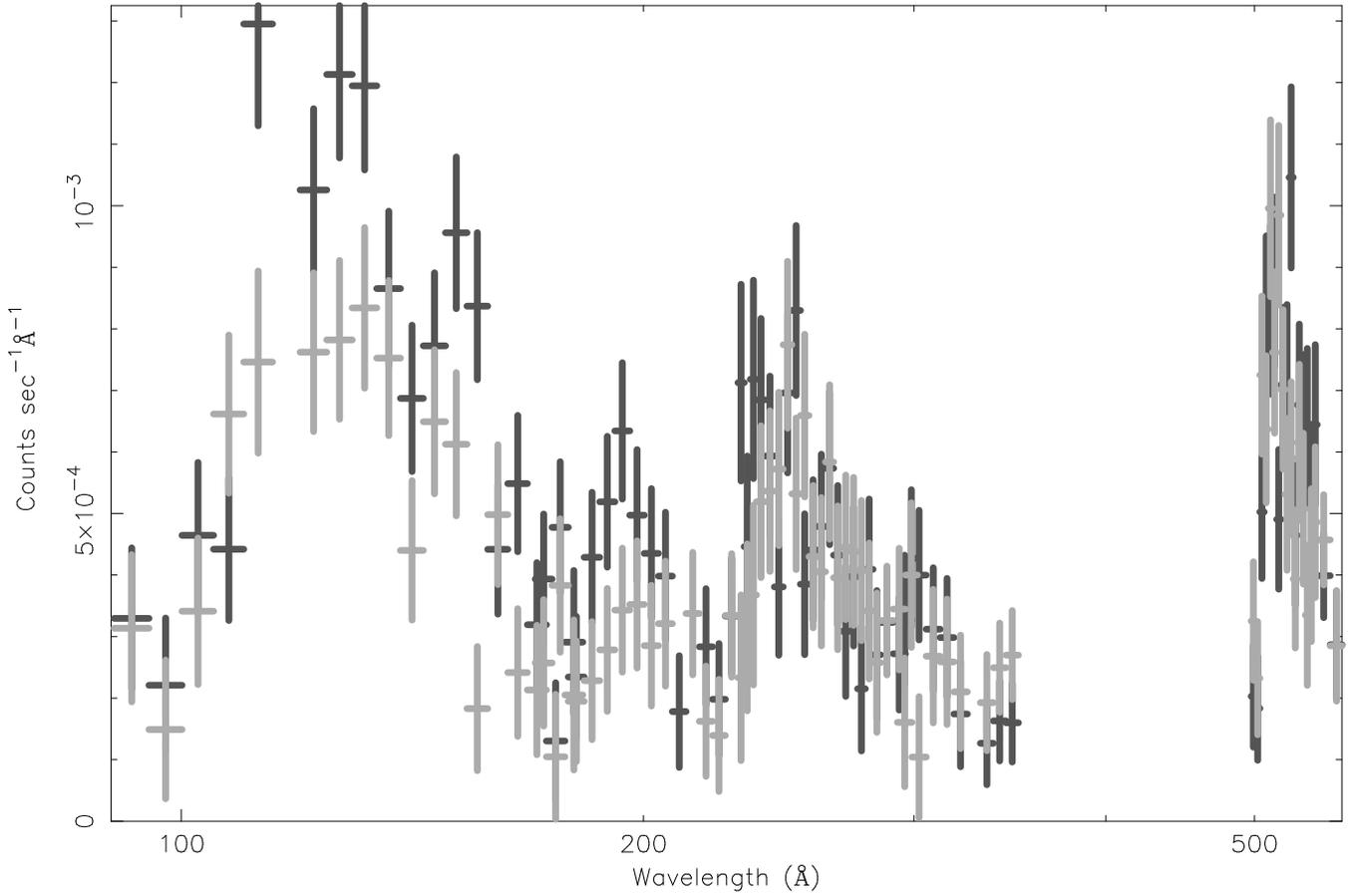}}
\caption{EUV spectra of \object{RE J0720-318} corresponding to phases
$\Phi = 0.4-0.6$ (grey) and $\Phi = 0.9-0.1$ (black). It can clearly be seen that the
change in EUV flux is confined to $\lambda < 240$\AA. This is consistent with
an apparent change in photospheric opacity levels.}
\label{Spectra}
\end{figure*}

Light-curves of \object{RE J0720-318} were extracted from the four
detectors onboard EUVE and were analysed with a Fourier
routine based on the Lomb algorithm (Lomb
\cite{lomb76}) to look for power
 between 0.3 -- 560.0 days$^{-1}$. The three spectrometer light-curve 
power spectra all have high noise
 levels and no significant periodicities other than
 those related to the satellite orbit were found. Nevertheless, we
noted a weak
feature in the SW power spectrum centred on $\sim 2.17$
 days$^{-1}$. In contrast, in the power spectrum of the DS light-curve
we found a significant peak at a confidence level greater
than 3$\sigma$, centered on 2.160
 days$^{-1}$. We note that the South Atlantic Anomaly may introduce false peaks into the power spectra of EUVE lightcurves, principally at 0.98 days (Halpern and Marshall \cite{halpern}), but also including an alias at half this cycle time. Power spectra of the primsching and deadtime correction tables for this observation do indeed reveal a periodicity on 0.989 days. However, the weak alias we detect at 0.50 days appears to be distinct from our 0.463 day period. Excluding satellite orbital power no other
notable features are present. The DS light-curve and relevant
section of the power spectrum are presented in Fig.~\ref{Period search}. 
Fitting a sinusoid to this light-curve, the epoch of maximum EUV count rate
($70-180$\AA) was found to be $\mathrm{T}_{0}\mathrm{(HJD)}
 = 2450068.677 \pm 0.006$. Folding the
four light-curves on this ephemeris it was apparent that while the SW flux is
also modulated on this period (Fig.~\ref{Folded}), within
statistical errors, both the MW and LW fluxes are constant in time, further suggesting that the 0.463 day period is real.  Accepting this, the flux change is likely to be due to either
variations in HeII opacity along the line of sight or the rotational
modulation of spatial inhomogeneities
in the photospheric helium and metal distribution. Two spectra
corresponding to phases $\Phi = 0.4-0.6$ and $\Phi = 0.9-0.1$ of the folded DS light-curve were
 extracted. These are shown in Fig.~\ref{Spectra}. It can be seen that the
variation in EUV flux actually occurs shortward of
$\lambda \sim 240$\AA. This observation is inconsistent
 with the line of sight explanation outlined above. In such a
case, any change would be restricted to $\lambda < 228$\AA. We suggest
therefore that the flux modulation is most likely to be due to a
$0.463 \pm 0.004$ day rotation of the white dwarf arising from inhomogeneous surface abundances. 
Such inhomogeneities on the
surfaces of the white dwarfs in pre-CV systems \object{V471 Tauri} (Jensen et
al. \cite{jensen86}, Dupuis et al. \cite{dupuis97b}) and \object{RE
J1016-053} (Vennes et al. \cite{vennes96b}) are believed to be the
signature of accretion from the wind of the secondary. 

\subsubsection{Phase Resolved EUV Spectra}
\begin{table}
\caption[Spectral Best Fit]{Parameter values for the 'best fitting' global
model to the EUVE data.}
\label{table4}
\small
\begin{center}
\begin{tabular}{cc} 
\hline 
Model Parameter & Value($\pm1\sigma$ ISM columns) \\ \hline
T & 55000K \\
log g & 7.50 \\
N(HI) &  $2.45(2.38-2.51) \cdot 10^{18} \mathrm{cm}^{-2}$ \\
N(HeI) &   $1.36(1.30-1.43) \cdot 10^{18} \mathrm{cm}^{-2}$ \\
N(HeII) &  $7.93(4.16-11.23) \cdot 10^{17} \mathrm{cm}^{-2}$ \\
log (He/H) & $-3.77$ \\
log (C/H) & $-5.41$  \\
log (N/H) & $-5.82$  \\
log (O/H) & $-5.21$ \\
log (Ne/H) & $-6.00$ \\
log (Si/H) & $-6.41$  \\
log (Fe/H) & $-6.91$ \\ \hline
\end{tabular}
\end{center}
\end{table}
To constrain the nature of the surface abundance inhomogeneities, the
two phase resolved spectra were examined with another small grid of
photospheric models. This grid, hereafter grid 1, was built around the model shown in Table~\ref{table4}
such that this model point defined the grid centre. 
Grid 1 is specified by two
independent parameters, the helium abundance, Y and the metal abundance, Z. The
metal abundance ratios are fixed with respect to each other and are identical to those of
the model described in Table~\ref{table4}. Metallicity, Z, can only vary
as a whole and Z is directly indexed to the model
carbon abundance. During the fitting procedure, two copies of model grid 1 were
simultaneously fit to the two phased spectra while holding the line of
sight columns fixed at the values given in Table~\ref{table4}. In this
procedure, the 
$\chi^{2}$-statistic provides a measure of the combined fit quality of
the models to
the two spectra. In a first
step, to test the
assumption that the spectral change was due to non-uniformities in the photospheric helium
distribution the two grid metal parameters were coupled to each other. The
best fit to the two spectra was found and the $\chi^{2}$ noted. 
Second, to test the assumption that the spectral
change was due to non-uniformities in the photospheric metal
distribution the grid helium parameters were coupled to each other. Again
the
best fit to the two spectra was found and the $\chi^{2}$
noted. The best fit model parameters for the two steps are given in Table~\ref{table5}, where
subscripts b and d refer to spectral phases $\phi = 0.9-0.1$ and
$\phi = 0.4-0.6$ respectively. As the two fitting steps have the same
number of degrees of freedom, we may use the
likelihood statistic of Edwards (\cite{edwards72}), given by the
formula $L_{1}/L_{2} = \exp[(\chi_{2}^{2}-\chi_{1}^{2})/2]$, 
to calculate that the spectral
changes are $\sim 100\times$ more likely to be the result of helium as
 opposed to metal inhomogeneities on the surface
of the white dwarf. Indeed, no significant improvement in the fit is achieved
by allowing the helium and metal parameters to vary completely
independently.
\begin{table}
\caption[Spectral Variations]{Best fit model parameters assuming, in
turn, that helium and metals are homogeneously spatially distributed across the
stellar photosphere. Y is the helium abundance, Z is the metal
abundance. For both fits $\nu = 177$.}
\label{table5}
\small
\begin{center}
\begin{tabular}{cc} \hline
Parameter & Value (log (N/H)) \\ \hline
Z$_{b}$        &  $-5.50(-5.53 - -5.44)$    \\
Y$_{b}$ = Y$_{d}$        &  $-3.70(-3.75 - -3.65)$    \\
Z$_{d}$        &  $-5.37(-5.42 - -5.32)$      \\
$\chi^{2}$     &  250.4     \\ \hline
Z$_{b}$ = Z$_{d}$       &  $-5.44(-5.50 - -5.39)$     \\
Y$_{b}$        &  $-3.78(-3.84 - -3.70)$     \\
Y$_{d}$        &  $-3.60(-3.66 - -3.54)$     \\ 
$\chi^{2}$     &  241.3    \\ \hline \hline
\end{tabular}
\end{center}
\end{table}

With this in mind we have examined
the possibly that the photospheric helium is confined to a limited region of the white dwarf
surface. This was carried out by attempting to
fit a linear combination of grid 1 and a further grid
made up of models
representative of the radiative levitation calculations of Chayer et
al (\cite{chayer95}), hereafter grid 2, and described in Table~\ref{table6}.  Grid 2 has only one variable
parameter, log (Fe/H). This may vary between an abundance of
$-5.2$, consistent with the theoretical prediction of Chayer et al. (\cite{chayer95}) and
a lower limit of $-7.0$. The abundances
of the other elements included, He,C,N,O,Si and Ne, are fixed at
 the values given in Table~\ref{table6}.  
\begin{table}
\caption[Radiative levitation model]{The parameters of the model
representative of the predictions of radiative levitation theory.}
\label{table6}
\small
\begin{center}
\begin{tabular}{cc} 
\hline 
Model Parameter & Value($\pm1 \sigma$) \\ \hline
T & 55000 K \\
log g & 7.50 \\
log (He/H) & $-5.00$ \\
log (C/H) & $-5.50$ \\
log (N/H) & $-5.50$ \\
log (O/H) & $-5.50$ \\
log (Ne/H) & $-6.20$ \\
log (Si/H) & $-7.20$ \\
log (Fe/H) & $-5.20 - -7.00$ \\ \hline
\end{tabular}
\end{center}
\end{table}
With the grid 1 Y fixed at the grid upper limit of $-3.15$,
the grid 1 Z fixed at $-5.5$, and with the
grid 2 iron abundance initially fixed at the theoretical value of
$-5.20$, by
allowing the ISM columns and relative model proportions to vary
freely, no acceptable fit to the integrated EUV spectrum could be obtained. The predicted
 SW flux was several orders of magnitude
too low. Next, the fitting procedure was repeated with a freely varying grid
2 iron abundance.  An accurate match to the EUV 
data  was found with log (Fe/H) = $-7.0$ and
relative model proportions of
$\sim$60\%\ and $\sim$40\%\ for grid 1 and grid 2 respectively. The
fit achieved here to the EUV data is as statistically
significant as that of the model described in Table~\ref{table4}. As this linear
combination is also as successful in reproducing
the UV observations as our single component model, one cannot be
preferred over the other. The EUV spectral variation may be
reproduced by a $\sim$20\%\ change in the relative proportions of the
models i.e. $\phi = 0.4-0.6$ -- 70\%\ grid 1 + 30\%\ grid 2 and
$\phi = 0.9-0.1$ -- 50\%\ grid 1 + 50\%\ grid 2. 
 
\section{Discussion}

\subsection{Is \object{RE J0720-318} viewed through a circumbinary disk$?$}

The two previous studies of the EUV spectrum of \object{RE J0720-318} both
suggested that the peculiar N(HI)/N(HeI) $\sim$ 1 along this line of
sight may be due to the presence of a circumbinary disk (Burleigh et
al. \cite{burleigh97}, Dupuis et al. \cite{dupuis97a}), a remnant of the prior phase of common
envelope evolution. In comparing the spectrum to that of the DA, RE
J0723-277, which lies at a distance of $\sim117$pc and is angularly
separated in sky from, the $\sim187$pc distant, \object{RE J0720-318} by
4${^\circ}$, Dupuis et al. (\cite{dupuis97a}) alternatively suggested that the
line of sight towards the DAO intercepts an ionized cloud. However,
neither study made a comparison between the columns observed towards
the white dwarfs and the B stars $\epsilon$CMa and $\beta$CMa, two
other objects in this region of sky which have been spectroscopically
observed by EUVE. Using a grid of LTE, homogeneous H+He photospheric models
calculated by Detlev Koester our measurements of the columns towards RE
J0723-277 are
N(HI)$=8.56-9.72\cdot10^{17}\mathrm{cm}^{-2}$,
N(HeI)$=1.38-1.60\cdot10^{17}\mathrm{cm}^{-2}$ and
N(HeII)$=0.64-1.98\cdot10^{17}\mathrm{cm}^{-2}$, consistent with the
values found by Dupuis et al. (\cite{dupuis97a}). The  
implied ionization fractions of f$_{H}=0.51-0.76$ and f$_{He}=0.28-0.59$ for H and He
respectively are consistent with the typical
values found by Barstow et al. (\cite{barstow97}) in a spectroscopic EUV
study of 13 DA white
dwarfs. It has been suggested that the similarities in ionization
fractions observed towards the white dwarfs in their sample may
indicate that the majority of the absorbing material lies within the
local interstellar cloud (LIC; Holberg et
al. \cite{holberg98b}). Indeed, Holberg et al. (\cite{holberg98b}) have
found strong evidence supporting this from a high resolution HST STIS
observation of the DA white dwarf RE J1032+53. Cassinelli et al. (\cite{cassinelli95}) report N(HI)$=7-12\cdot10^{17}\mathrm{cm}^{-2}$ towards
$\epsilon$CMa, which Hipparcos (ESA. \cite{hipparcos}) measurements put at a distance of
between 123-143pc. This is very similar to the measurement of N(HI)
towards \object{RE J0723-277} obtained here. Although constraints on N(HeI)
towards $\epsilon$CMa are not specifically given, examination of
Fig. 5b of Cassinelli et al. (\cite{cassinelli95}) suggests that the neutral helium
column along this line of sight may also be consistent with that
observed towards the DA ($\sim10^{17}\mathrm{cm}^{-2}$). In fact
high resolution observations of the line of sight towards
$\epsilon$CMa obtained with the HST echelle B and G190M gratings,
indicate that most of the intervening material lies within 3pc of the
Sun, consistent with residence in the LIC (Gry et al. \cite{gry95}).
Over the remaining distance, the mean volume density is found to be less
than $2\cdot10^{-4}\mathrm{cm}^{-3}$. It seems likely then, that the opacity
observed towards \object{RE J0723-277} also resides within the LIC. In stark
contrast, the columns observed towards $\beta$CMa, only 14$^{\circ}$
away in the sky from $\epsilon$CMa, but more distant at between
139-170pc, are  N(HI)$=2.0-2.2\cdot10^{18}\mathrm{cm}^{-2}$ and
N(HeI)$>1.4\cdot10^{18}\mathrm{cm}^{-2}$ (Cassinelli et al. \cite{cassinelli96}). This implies a ratio of N(HI)/N(HeI)$\simless1$ along this
line of sight, very similar to that found towards \object{RE J0720-318}.  A
comparison between the column densities measured towards these two
objects, bearing in mind N(HI)=$2.38-2.51\cdot10^{18}\mathrm{cm}^{-2}$
and N(HeI)$=1.30-1.43\cdot10^{18}\mathrm{cm}^{-2}$ towards \object{RE
J0720-318}, may in fact indicate that both $\beta$CMa and the DAO are
viewed through the same LISM structure. The existence of an ionized
gas cloud lying beyond the local cloud and in the direction of these
objects is strongly supported by the findings of a high resolution
study of the line of sight towards $\beta$CMa (Dupin \& Gry \cite{dupin98}). It
appears that this is dominated by two ionized clouds distinct from
each other in velocity space by 10kms$^{-1}$, and which Dupin \& Gry refer to as C and D. Neither has a velocity consistent with the LIC. Using sulphur, which is only slightly depleted in the interstellar environment, to map the distribution of hydrogen gas, they calculate that the total hydrogen column along this line is N(H$_{\mathrm{tot}}$)$=1.7-2.1\cdot10^{19}\mathrm{cm}^{-2}$. This is approximately $10\times$ the neutral helium column observed towards \object{RE J0720-318}. If the interpretation here, that these objects are viewed through the same structure, is correct, then such a region of ionized gas, residing at between 123-170pc away in the direction of the CMa tunnel, extends for more than 40pc in length. Further STIS observations of \object{RE J0720-318} and RE J0723-277 would be able to confirm these conclusions. Additionally, a long duration EUVE observation of \object{RE J0720-318} might allow          
a more accurate determination of f$_{He}$ along this line of sight which may,
in turn, shed further light on the mechanisms responsible for the observed degree of He
ionization. For example, consistency with the fractions
found by Barstow et al. (\cite{barstow97}) may indicate that the
mechanism responsible for the level of f$_{He}$, operates over distances of
the order $\sim100$pc. This might pressumably argue against a model in
which the He ionization resulted from the ambient stellar EUV radiation field 
impinging on these clouds. 

\subsection{The elemental abundances in the atmosphere of an accreting
white dwarf}

In the Burleigh et al. (\cite{burleigh97}) analysis of \object{RE J0720-318},  LTE H+He models were used without the constraints imposed by the
 GHRS data. While the EUV spectrum was interpreted as originating from a white dwarf with a thin hydrogen 
layer of $3 \cdot 10^{-14}$M$_{\sun}$, a product of mass transfer during a prior common envelope phase, it was suggested that the homogeneously mixed helium observed in the visible photosphere was accreted from the red dwarf wind. However, the
 evidence from the light-curves, the phase resolved spectra and the
 multi-wavelength non-LTE modelling reported here, supports
 a wind accretion origin for all of the observed helium, from the optical through to the EUV. To first order, this excessive abundance of helium, $\simgreat$1 dex above the radiative levitation prediction, is maintained by a 
balance between the accretion of helium into and the gravitational settling of helium
out of the white dwarf photosphere and it is possible to obtain a
 crude estimate of the accretion rate by solving equation 1 (Sion \&
 Starrfield \cite{sion84}). However, such an estimate may be rather
 inaccurate for several reasons. For instance, in our calculation it
 is assumed that the relative abundance of elements in the incoming
 material have solar values, although it is possible that a slowly rotating weakly magnetic hot white dwarf 
preferentially accretes certain elements (Alcock \& Illarionov
 \cite{alcock80}). Furthermore, the size of the area of the photosphere onto which the material is accreted is unknown. Nevertheless, noting the following, it should be possible to eliminate the latter difficulty and obtain a rough lower
limit to the accretion rate. The spectroscopic signature of a given abundance of helium in the photosphere may be reproduced by models making various assumptions about the abundance and distribution of this helium. For example, a uniform distribution can be assumed over the entire area of the visible hemisphere, or a higher abundance can be assumed over a smaller region, as was shown earlier. In fact, it is found from modelling and equation 1 , where $\dot{\mathrm{M}}_{\mathrm{a}}$ is the mass accretion
rate, A, the accretion surface area, $\rho$ the mass density at formation 
depth of the HeII line used to measure the abundance and v
 the diffusion velocity relative to the centre of mass, also at the
line formation depth, 
 that the minimum accretion rate occurs for a uniform distribution of helium. 

\begin{equation}
\frac{\mathrm{(He/H)}_{\mathrm{\star}}}{\mathrm {(He/H)}_{\mathrm{\odot}}} = 
\frac{\dot{\mathrm{M}}_{\mathrm{a}}}{\dot{\mathrm{M}}_{\mathrm{a}} + \mathrm{A}\rho{\mathrm{v}}}
\end{equation}

The diffusion velocity has been calculated from equation 1 of Vennes
et al. 
(\cite{vennes88}),
 by incorporating a diffusion coefficient determined from the tables of
 Paquette et al. (\cite{paquette86}). Putting the following values into equation
1 , A$\approx 1.1\cdot10^{19}\mathrm{cm}^{2}$, $\rho \approx 8.5 \cdot 10^{-9}\mathrm{g cm}^{-3}$, 
log (He/H)$_{\star}\approx -3.9$ and $\mathrm{v} \approx 0.26 
\mathrm{cm s}^{-1}$ a lower limit of $\dot{\mathrm{M}}_{a} \simgreat 1 \cdot 10^{-18}$M$_{\sun}$ yr$^{-1}$ is determined.
 In fact optical data taken around 1995 January-February suggests that \object{RE J0720-318} undergoes periods of 
more intense accretion (Finley et al. \cite{finley97}). 
A H-layer mass of 
only $10^{-13}$M$_{\sun}$\ is
capable of hiding traces of the underlying helium envelope in a 60000K
hydrogen rich white dwarf (Vennes \& Fontaine \cite{vennes92}), and it is found that this DAO may have acquired 
$\sim 1.3 \cdot 10^{-12}$M$_{\sun}$\ of hydrogen since it 
emerged from the CE some 
$2.6 \cdot 10^{6}$ years ago. We conclude, that although \object{RE J0720-318} most likely has a hydrogen layer mass greater than $10^{-13}$M$_{\sun}$\ now, it may have been almost completely stripped of a hydrogen layer when it emerged from
the CE phase of evolution as initially proposed by Burleigh et al. (\cite{burleigh97}).

Both successful model fits to the EUV spectrum of \object{RE J0720-318} achieved here require a global iron abundance of log (Fe/H)$\sim -7.0$, although due to the limited wavelength coverage
 of the high resolution UV data it is not possible to exclude the likelihood that larger amounts of iron are in fact present in the atmosphere, and the
abundance of the other elements such as O and N is lower. However,
by assuming that an edge or line blend would be visible if predicted
to be larger than twice the S/N of the data,
the EUV spectrum can be used to set an upper limit of log
(Fe/H) $\simless -6.2$ on the iron abundance.  In the atmosphere of an
accreting white dwarf such as \object{RE J0720-318}, where the elemental
diffusion time-scales are very much
shorter than the evolutionary time-scale, the elemental abundances
should reach an equilibrium somewhere between those of the incoming
material and the radiative levitation values (e.g Vennes et al. \cite{vennes96a}). The theoretical radiative
levitation prediction (Chayer et al. \cite{chayer95})
which is highly sensitive to the effective temperature and surface
gravity of a star, for the abundances of iron and carbon is log (C/H)
$= -5.5\pm0.1$ and log (Fe/H) $= -5.1\pm0.1$. Again assuming the abundance of
elements in the accreting material
to be roughly solar it would be expected that the observed global iron
abundance should be at least log (Fe/H) $\sim -5.2$. In contrast to
the observed log (C/H), the measured iron abundance for \object{RE J0720-318} does not 
conform to a simple equilibrium prescription. Such inconsistencies between
the theoretical and observed abundances of elements have been widely
reported in previous studies of hot white dwarfs. For example
 Barstow et al. (\cite{barstow96})  find a silicon abundance in \object{GD394} at
least a factor 10 above that expected for a 38\,500K DA white dwarf.
In contrast, the silicon abundance of log (Si/H) $ = -7.98 \pm 0.07$ in \object{RE
J1614-085}, a DA of similar $T_{\mathrm{eff}}$, is around a factor 10 below the 
theoretical value, while nitrogen is found to be over abundant by
a factor 1000 (Holberg et al. \cite{holberg97}). Holberg et al. (\cite{holberg97})  suggest that
 both accretion and mass-loss are perhaps
responsible for the non-equilibrium abundances observed in RE
J1614-085. A mass-loss of only $10^{-16}$M$_{\sun}$ yr$^{-1}$ has been shown to be capable
of depleting the photosphere of a white dwarf of silicon on a time-scale of several
thousand years (Chayer et al. \cite{chayer97}). In fact a star of RE
J0720-318's metallicity (1/100th solar) is expected to
undergo mass-loss at a rate of $\sim 10^{-14}$M$_{\sun}$ yr$^{-1}$ (Abbott \cite{abbott82}). However, any mass-loss would also reduce the efficiency of
accretion (Mullan et al. \cite{mullan92}). The Bondi-Hoyle formalism can be used to provide a
theoretical estimate of the rate of spherically symmetric accretion of wind material. The prediction is extremely
sensitive to the relative
velocity of the incoming material, being inversely proportional to its fourth
power. If the separation between the primary and secondary binary
components is taken to be $\sim 4 \cdot 10^{11}$ cm, based on the ephemeris of
Vennes \& Thorstensen (\cite{vennes96a}), and conservative
estimates for the red dwarf mass-loss rate and relative wind velocity of
3 $ \cdot 10^{-14}$M$_{\sun}$ yr$^{-1}$ and $\sim 550$km s$^{-1}$
respectively (Mullan \cite{mullan96}) are adopted, an accretion rate of
$\sim 1 \cdot 10^{-16}$M$_{\sun}$ yr$^{-1}$ is predicted. Although the
 size of the area of the photosphere accreting is not known, detailed non-LTE modelling of the
1640\AA\ line profile with linear combinations of pure hydrogen and
mixed H+He composition synthetic spectra, allowing the helium
abundance and relative model proportions to vary freely, suggests that it is sizable. In fact if the helium
abundance is increased above log (He/H)$>-2.1$ corresponding to $\simless
35$\%\ of the visible surface, then a deterioration in the
goodness of fit occurs
at the 3$\sigma$ level. A solution of equation 1 for these parameters
has $\dot{\mathrm{M}}_{\mathrm{a}}\simless3\cdot10^{-17}{\mathrm{M}}_{\sun}{\mathrm{yr}}^{-1}$.
 We stress that this estimate is extremely crude. However, in view of the observed upper limit on the iron abundance we suggest that, while
material is being accreted into the atmosphere of this white dwarf, a weak radiatively driven wind may selectively expel heavy elements from the photosphere. 

A EUV study of another DAO+dM system, RE J1016-053, detected in the WFC
survey  (Pounds et al. \cite{pounds93}) and having
a binary period of 0.789 days (Thorstensen et al. \cite{thorstensen96}) also found low globally averaged heavy element abundances on the white dwarf. The line of sight neutral hydrogen column towards this binary of
$2.65 \cdot 10^{19}\mathrm{cm}^{-2}$ (Vennes et al. \cite{vennes96b}) severely attentuates the EUV spectrum longward of 180\AA\ making it more or less
impossible to differentiate between helium or metal spatial
abundance inhomogeneities as responsible for the 30\%\ peak to peak
variation on a 57.3 minute period observed in the EUV flux. The low global
heavy element abundances led  Vennes et al. (\cite{vennes96b}) to suggest that perhaps metals were confined to a small region of the photosphere and were responsible for this variation. 
However, modeling indicates that the spectral changes observed in RE J1016-053 can be reproduced by an apparent global change in helium abundance of
$\sim 0.15$ dex, similar to the result obtained here for \object{RE J0720-318}. It seems more plausible that the observed spectral changes are due mainly to inhomogenieties in the spatial distribution of photospheric helium.

\subsection{Where do the weak red-shifted components originate$?$}
Weak, non-photospheric, high ionization features such as the
resonance lines of CIV, SiIV and NV have been reported in the UV
spectra of several white dwarfs (e.g. Shipman et
al. \cite{shipmann95}; Holberg et al. \cite{holberg95}). These are often
observed to be blue shifted with respect to the photospheric
velocity and hence are frequently attributed to a shell of photoionized
circumstellar gas possibly
formed as a result of weak mass-loss. However, the weak CIV
lines detected in the GHRS spectra of \object{RE J0720-318} are redshifted by
$\sim50$km s$^{-1}$
with respect to the observed photospheric velocity, so this material must be moving towards the white dwarf, arguing against a circumstellar
origin. 

The close proximity of $\epsilon$ CMa and $\beta$
CMa, separated in the sky from \object{RE J0720-318} by $\sim6^{\circ}$ and
$\sim20^{\circ}$ respectively allows an examination of the possibility
that these features reside in the LISM. Distances to both stars are 
accurately known
from Hipparcos (ESA. \cite{hipparcos}) observations, $\epsilon$ CMa lies between 123-143pc
away and $\beta$
CMa between 139-170pc. In their high
resolution study of $\epsilon$ CMa with the HST, Gry et al. (\cite{gry95}) 
tentatively detect two separate non-photospheric components to the CIV
resonance line profiles. The first, with a column density of
 $4.2 \pm 1.0 \cdot 10^{12}$cm$^{-2}$ and velocity of
$-10\pm2\mathrm{kms}^{-1}$ they associate with a warm cloud probably
lying beyond the LIC. The other, with a column desnity of $3.2 \pm 1.0 \cdot 10^{12}$cm$^{-2}$ has a velocity of
$17\pm2\mathrm{kms}^{-1}$ which is consistent with the projected
velocity of the LIC along this line of
sight. The velocity and strength of the possible
non-photospheric CIV resonance lines observed in spectrum of $\beta$
CMa are also consistent with a formation in the LIC (Dupin \& Gry
\cite{dupin98}). However, for the CIV features in the GHRS spectra of RE
J0720-318, although the weighted mean velocity of $17.1 \pm
2.9$kms$^{-1}$ suggests a formation in the LIC, assuming the lines to lie on the linear part
of the curve-of-growth, a lower limit to the column density of
N(CIV)$\simgreat1\cdot10^{13}\mathrm{cm}^{-2}$ can be set. This may be
rather higher than expected from the $\epsilon$ and $\beta$CMa observational
results and the predictions of theoretical studies of the conductive
interface where this species is believed to reside (N(CIV)$\approx 3.1 \pm 1.0 \cdot 10^{11}$cm$^{-2}$; Cheng and
Bruhweiler \cite{cheng90}).
  
Alternatively,  it may be that they are related to
the binary itself. Indeed, the radiation field from a hot hydrogen rich white dwarf
 is capable of creating and maintaining
highly ionized species in a gas in close enough proximity (Dupree \&
Raymond \cite{dupree83}). For example, non-photospheric CIV absorption lines
have been observed at a wide range of binary phases in IUE echelle
 spectra of the DA+dM pre-CV binary
Feige 24 (Dupree \& Raymond \cite{dupree82}). Vennes \& Thorstensen (\cite{vennes94b}) used a simple photoionization
 model in
their study of low density gas in the environment of the DA, and found for plausible electron
temperatures in the range $10^{4}-10^{5}\mathrm{K}$, that while the majority of the
total carbon would be in the form of CV,  a fraction of between $10^{-2} - 10^{-4}$ would be in the form of CIV in its groundstate. They concluded that
a likely origin for the absorption features was in 
circumstellar material possibly a remnant left over from the CE phase
(Vennes and Thorstensen \cite{vennes94b}). As the HST observations of RE
J0720-318 were aquired close to binary quadrature, the
possibility that they form in a cloud of remnant CE gas located at the L1
Lagrangian point can be discounted. Nevertheless, using the
photoionization code XSTAR (Kallman \& Krolik \cite{kallman97}) it is found
that for gas with a plausible T$\sim10^{4}\mathrm{K}$, the elements of
which are in solar proportions, lying 
outside the binary system
at a distance of $\sim1000\times$ the binary
separation, where effectively
all CIV is in groundstate, the fraction of the total carbon in the CIV state,
 can be
 between $1\cdot10^{-4}-0.5$, for hydrogen densities in the range
$0.1-10^{4}\mathrm{cm}^{-3}$. In contrast, the fraction of helium in the neutral state is $\simless1\cdot10^{-12}$, with the majority being twice ionized. Thus it seems plausible that our CIV features may also 
originate in material lying outside of the binary system, possibly
remnants of the CE. Indeed, as pointed out by both Burleigh et
al. (\cite{burleigh97}) and Dupuis et al. (\cite{dupuis97a}), a  
circumbinary disk is theoretically predicted to form from the remains
 of the common envelope gas. 

 Further modelling of gas in the vicinity of the
DAO was carried out to investigate whether these features may alternatively
reside in a wind from the red dwarf companion. Departures from
groundstate are likely negligible at the densities and
temperature considered (Kallman \& Krolik \cite{kallman97}), except within a few stellar radii of the white
dwarf where photoexcitation may be significant. XSTAR was run to
calculate the fraction of total gas  
in the form of CIV for a range of mass-loss
rates ($10^{-9}-10^{-15}{\mathrm{M}}_{\sun}{\mathrm{yr}}^{-1}$). The
wind density was assumed to
drop off as the square of the distance from the source and the wind 
velocity taken as $\sim100\mathrm{kms}^{-1}$, roughly the
difference between the radial velocity of the CIV lines and the ephemeris prediction
for the red-dwarf velocity. At a series of logarithmically spaced distances from the white
dwarf, the local number density of CIV was calculated. By integrating along the line of sight an
estimate for the CIV 
column was obtained. For the chosen mass-loss rates the predicted column was found to lie in the range N(CIV)$\approx1\cdot10^{17}-3\cdot10^{7}\mathrm{cm}^{-3}$. A mass-loss
rate of $10^{-11}{\mathrm{M}}_{\sun}{\mathrm{yr}}^{-1}$ should produce
N(CIV)$\approx1.7\cdot10^{13}\mathrm{cm}^{-3}$ consistent with the
observational lower limit. Mullan (\cite{mullan96})  reports that a detection
of the inactive Barnard's Star at $\lambda=6\mathrm{cm}$ has been interpreted as free-free emission originating in a mass-loss wind of
$2\cdot10^{-11}{\mathrm{M}}_{\sun}{\mathrm{yr}}^{-1}$. Hence, it also seems
plausible that these absorption features may be formed in a
photoionized wind from the red dwarf. Furthermore, at the observed
velocity, wind material would be moving towards the white dwarf and
possibly accreting onto it, consistent with the wind accretion model 
cited to explain the EUVE data.
  
\section{Summary}

It has been demonstrated that {\it all} of the helium in the optical,
UV and EUV spectra of \object{RE J0720-318}, and most probably heavy
elements too, are likely accreted into
the atmosphere from the wind of the dM companion. This conclusion is supported by the non-uniform
 spatial distribution of EUV opacity in
the white dwarf photosphere.  It has been found that even at
the lower limiting rate of accretion, the white dwarf may well have gained
its entire observable atmosphere from the wind of the secondary in the
time since it emerged from the CE phase of evolution. Furthermore, the
non-uniformity has allowed an estimate of
the rotation period of the white dwarf ($0.463\pm0.004$days) to be made. The
photospheric iron abundance is at least a factor 10 lower than predicted
by a simple balance between gravitational settling and the opposing
upward pressure of the
radiation field. Along with the rather lower than expected
accretion rate, this has been interpreted as due to the possible presence of
a selective mass-loss wind. Several possible 
origins for weak non-photospheric CIV lines
observed in GHRS spectra of \object{RE J0720-318} have been
examined. Although the line velocities are consistent with a  
formation in the
LISM, the column density may be rather larger than
expected. Alternatively, we have shown that this CIV may reside in 
photoionized gas in close proximity to the white dwarf. This may take
the form of either residual CE material surrounding the binary or in a
putative wind from the secondary star. Finally, circumstantial evidence has been
 presented which argues against a 
circumbinary disk as responsible for the unusual N(HI)/N(HeI)$\sim1$ along
this line of sight. Instead, it has been shown that it is  more likely
due to an
extended region of ionized gas lying in the direction of the CMa ISM
tunnel.

\begin{acknowledgements}
PDD, MAB and MRB acknowledge the support of PPARC, UK.
\end{acknowledgements}

\end{document}